\documentclass[conference]{IEEEtran}
\IEEEoverridecommandlockouts
\usepackage{cite}
\usepackage{amsmath,amssymb,amsfonts}
\usepackage{multirow}
\usepackage{array}
\usepackage{mathtools}
\usepackage{algorithmic}
\usepackage{graphicx}
\usepackage{textcomp}
\usepackage{xcolor}

\newcolumntype{P}[1]{>{\centering\arraybackslash}p{#1}}
\newcolumntype{M}[1]{>{\centering\arraybackslash}m{#1}}
\def\BibTeX{{\rm B\kern-.05em{\sc i\kern-.025em b}\kern-.08em
    T\kern-.1667em\lower.7ex\hbox{E}\kern-.125emX}}
\begin{document}

\title{Is my Neural Network Neuromorphic? Taxonomy, Recent Trends and Future Directions in Neuromorphic Engineering
}
\author{\IEEEauthorblockN{Sumon Kumar Bose}
\IEEEauthorblockA{\textit{School of EEE} \\
\textit{Nanyang Technological University}\\
}
\and
\IEEEauthorblockN{Jyotibdha Acharya}
\IEEEauthorblockA{\textit{School of EEE} \\
\textit{Nanyang Technological University}\\
}
\and
\IEEEauthorblockN{Arindam Basu}
\IEEEauthorblockA{\textit{School of EEE} \\
\textit{Nanyang Technological University}\\}
}

\maketitle
\begin{abstract}
In this paper, we review recent work published over the last 3 years under the umbrella of Neuromorphic engineering to analyze what are the common features among such systems. We see that there is no clear consensus but each system has one or more of the following features:(1) Analog computing (2)  Non von-Neumann Architecture and low-precision digital processing (3) Spiking Neural Networks (SNN) with components closely related to biology. We compare recent machine learning accelerator chips to show that indeed analog processing and reduced bit precision architectures have best throughput, energy and area efficiencies. However, pure digital architectures can also achieve quite high efficiencies by just adopting a non von-Neumann architecture. Given the design automation tools for digital hardware design, it raises a question on the likelihood of adoption of analog processing in the near future for industrial designs. Next, we argue about the importance of defining standards and choosing proper benchmarks for the progress of  neuromorphic system designs and propose some desired characteristics of such benchmarks. Finally, we show brain-machine interfaces as a potential task that fulfils all the criteria of such benchmarks. 
\end{abstract}

\section{Introduction}
\label{sec:intro}
The rapid progress of Machine Learning (ML) fuelled by Deep Neural Networks (DNN) in the last several years has created an impact in a wide variety of fields ranging from computer vision, speech analysis, natural language processing etc. With the progress in software, there has been a concomitant push to develop better hardware architectures to support the deployment as well as training of these algorithms\cite{jetcas_review,jetcas_boris}. This has rekindled an interest in ``Neuromorphic Engineering"--a term coined in 1990 by Carver Mead in his seminal paper \cite{mead1990} where he claimed that hardware implementations of algorithms like pattern recognition (where relative values are of more importance than absolute ones e.g. is this image more likely to be a cat or a dog?) would be more energy and area efficient if it adopts biological strategies of analog processing.S

\begin{figure}[h!]
\centering
\includegraphics[scale=0.37]{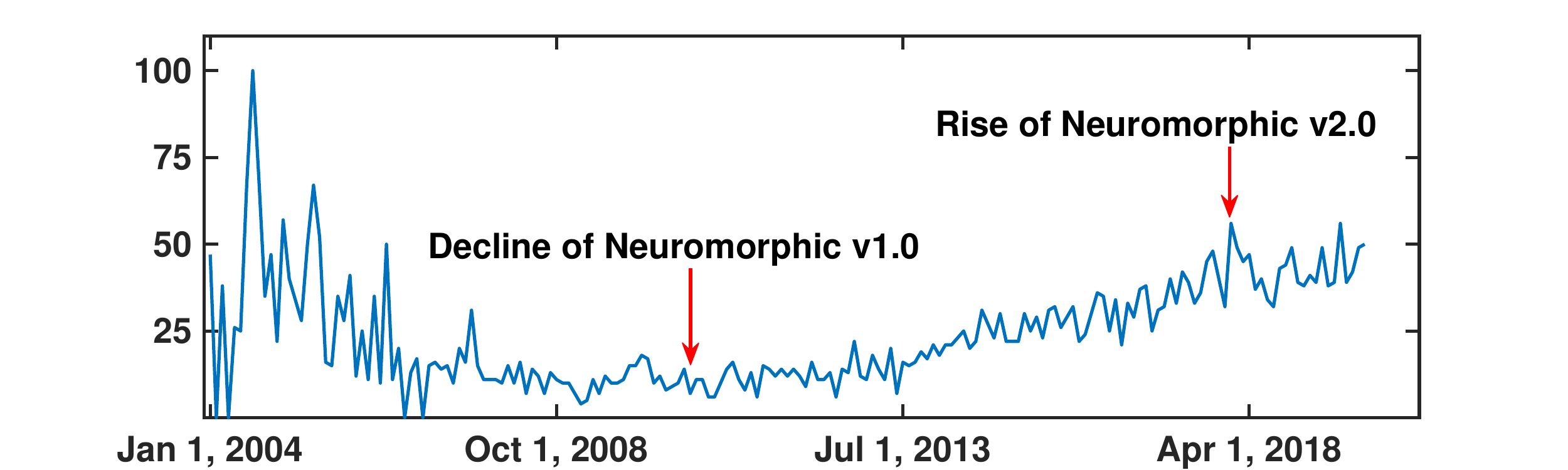}\\
(a)\\
\includegraphics[scale=0.37]{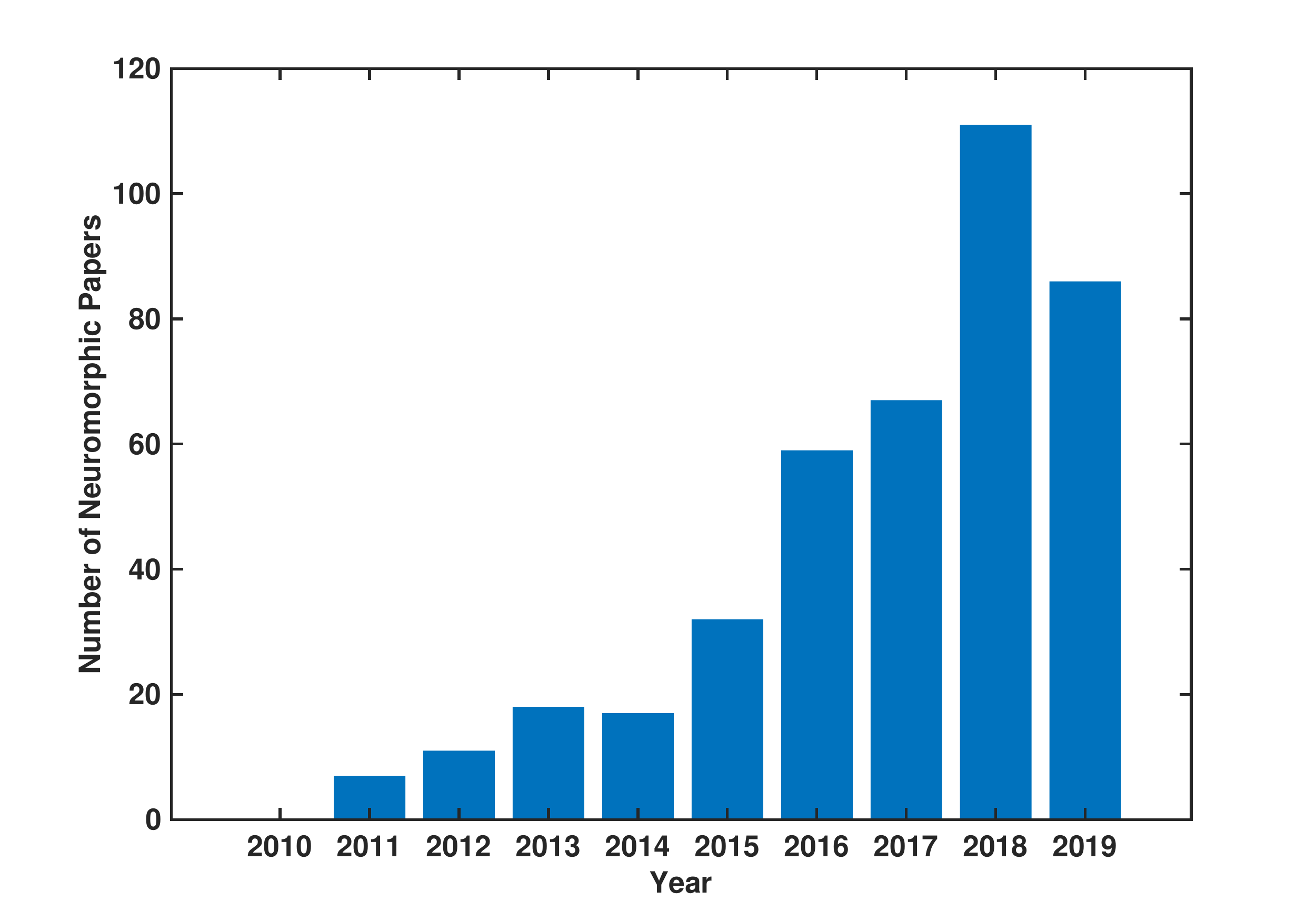}\\
(b)
\caption{(a) Google search trends over the last 15 years for the topic ``Neuromorphic Engineering" shows a decline around 2010 followed by a renewed interest in the last 5 years. (b) Number of neuromorphic papers published in journals from the Nature series have shown a steady increase in the last 10 years. Data for 2019 is till the month of October. }
\label{fig:googletrend}
\end{figure}

While the above idea of brain-inspired analog processing is very appealing and showed initial promise with several interesting sensory prototypes, it failed to gain increased traction over time possibly due to the potential difficulties of creating robust, programmable, large-scale analog designs that can benefit from technology scaling in an easy manner. However, in the last 5 years, there has been renewed interest in this topic, albeit with a slightly expanded connotation of the term ``neuromorphic". Figure \ref{fig:googletrend}(a) shows a history of google searches of the term ``neuromorphic engineering" over the past 15 years (obtainable from Google Trends). Data points are plotted for every month with the maximum search number normalized to $100$. It can be seen that there was a decline in interest about neuromorphic research around 2010. However, it has again gained momentum in the last five years with a slightly broadened scope which we refer to as version 2 (while referring to the Meadian definition as version 1). A similar trend (plotted in Figure \ref{fig:googletrend}(b)) is obtained also by analyzing the number of papers published in relevant journals (Nature, Nature Communications, Nature Electronics, Nature Machine Intelligence, Nature Materials, Nature Nanotechnology) from the Nature journal series over the last $\approx 10$  years that are on the topic of neuromorphic research. It can be seen that there is a rapid increase in the number of such papers over the last 5 years.

The rest of the paper is organized as follows: the next section introduces the new connotation of the term ``neuromorphic" followed by an analysis of some recent research trends in this field. Section \ref{sec:benchmarks} describes the need for neuromorphic benchmarks and some desired criteria of such benchmarks while Section \ref{sec:bmi} proposes brain-machine interfaces as a potentially good benchmark. 
%\begin{figure}[h!]
%\centering
%\includegraphics[scale=0.37]{image/number_n_papers.eps}
%\caption{Number of neuromorphic papers}
%\label{fig1}
%\end{figure}
\section{Neuromorphic v2.0: A Taxonomy}
\label{sec:taxonomy}
%\begin{figure}[h!]
%\centering
%\includegraphics[scale=0.4]{image/vnv_architecture.eps}
%\caption{Illustration of von-Neumann (left) and non von-Neumann (right) architectures. While a common shared memory is used in the former, the latter has a distributed memory interspersed within processing elements.}
%\label{fig:von-neumann}
%\end{figure}
As discussed in the last section, the renaissance in Neuromorphic research over the last 5 years has seen the term being used in a wider sense than the original definition\cite{mead1990}. This is partially due to the fact that scientists from different communities (not only circuits or neuroscientists) ranging from material science to computer architects have now become involved. Based on the recent work, we describe next the key characteristic features of this new version of neuromorphic systems as:
\begin{itemize}
    \item Use of \textbf{analog or physics based processing} as opposed to conventional digital circuits--this is same as the original version of neuromorphic system from a circuits perspective.
    \item From the viewpoint of computer architecture, usage of \textbf{non von-Neumann architecture} (independent of analog or digital compute) and \textbf{low-precision digital datapath} are hallmarks of neuromorphic systems. In other words, conventional computers using von-Neumann architectures read from memory, compute and write back the result--this is very different from brain-inspired systems where memory and computing are interspersed \cite{liu_indiveri}.
    \item Computer scientists and algorithm developers on the other hand consider a system neuromorphic if it uses a \textbf{spiking neural network} (SNN) as opposed to a traditional artificial neural network (ANN). Neurons in an SNN inherently encode time and output a 1-bit digital pulse called a spike or action potential.
\end{itemize}
\begin{figure}[h!]
\centering
\includegraphics[scale=0.30]{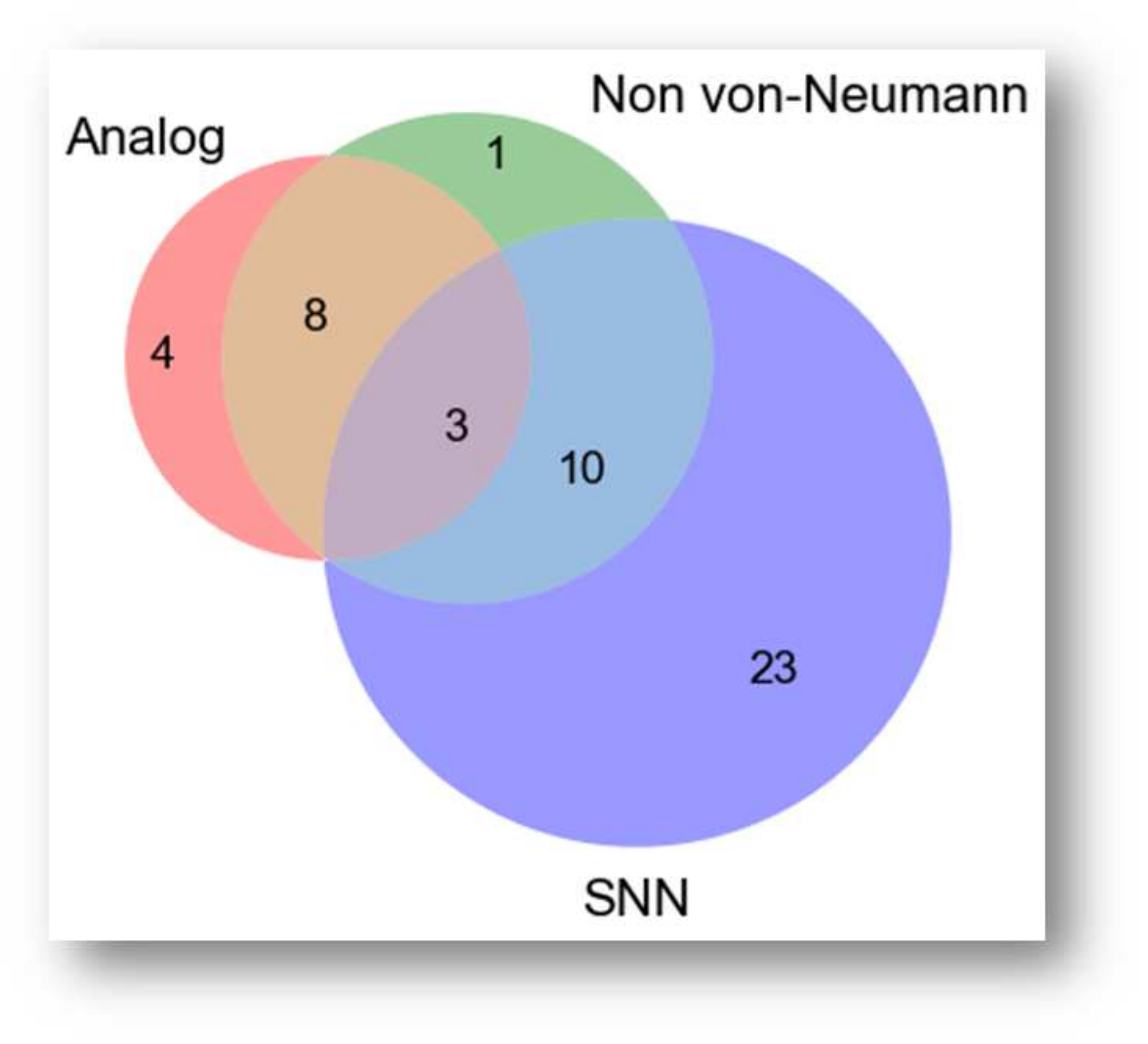}
\caption{Survey of neuromorphic systems reported over 2017-2019 in  Nature, Science, Science Advances, Nature Nanotechnology, Nature Electronics, Nature Materials, Nature Communications
%~\cite{pei2019towards}~-\cite{wu2018skin}
. A large majority use SNN in their work. Details of all papers used in the survey are in \cite{Survey_excel}.}
\label{fig:venn_science}
\end{figure}
\begin{figure}[h!]
\centering
\includegraphics[scale=0.30]{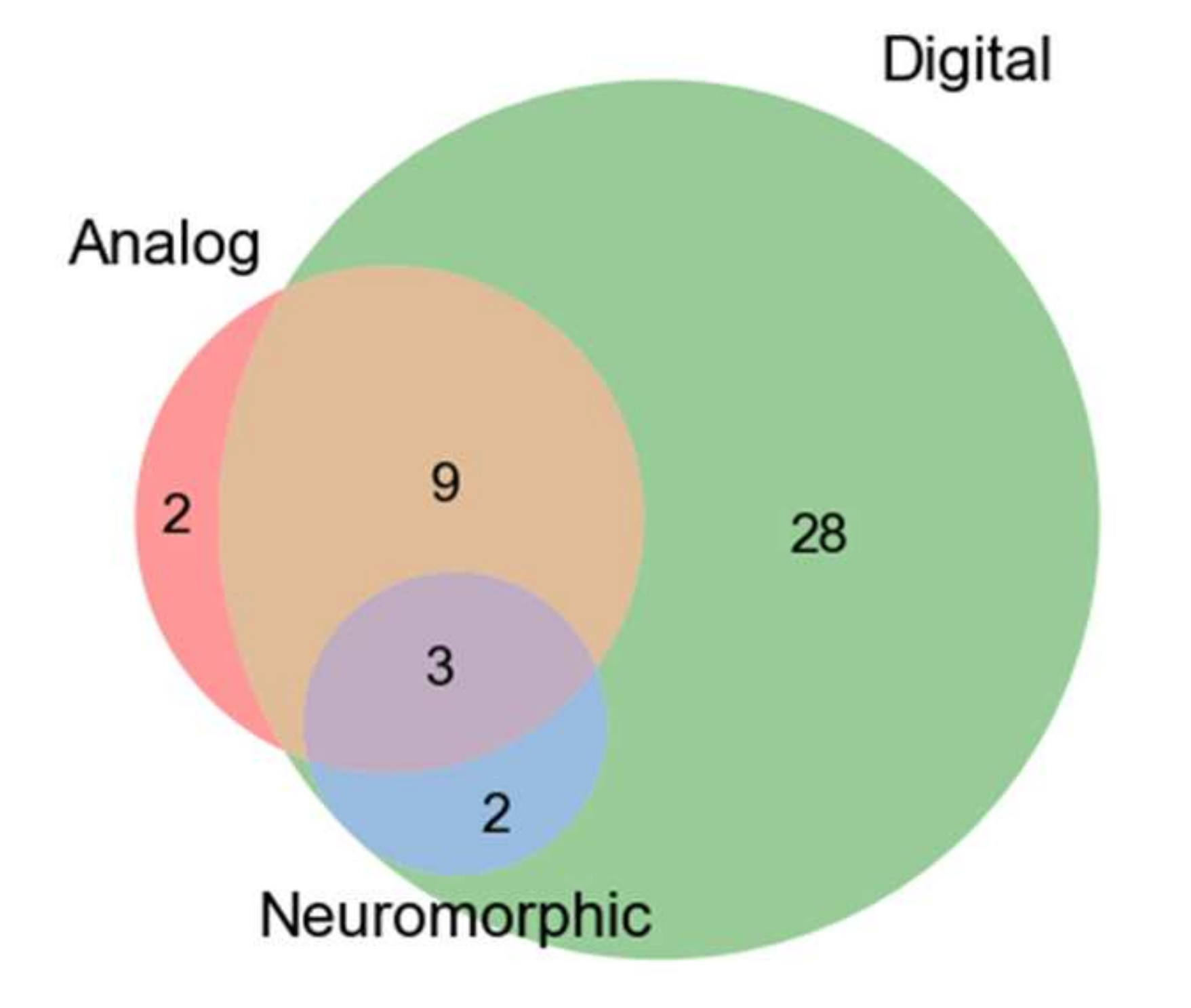}
\caption{Survey of IC implementations of non von-Neumann architecture over the same period in ISSCC, SOVC, JSSC
%~\cite{A_29TOPS}~-\cite{sovc_2019_6} 
however shows very few work uses the term ``neuromorphic". Details of all papers used in the survey are in \cite{Survey_excel}.}
\label{fig:venn_asic}
\end{figure}
We next illustrate how frequently each type of viewpoint is expressed in neuromorphic research. Figure \ref{fig:venn_science} categorizes the neuromorphic research papers published between 2017-2019 in the Nature series of journals surveyed in Figure \ref{fig:googletrend}(b) along with the journals Science and Science Advances. The papers are categorized according to the neuromorphic aspect they primarily focus on--(1) Analog processing, (2) non von-Neumann architecture or (3) SNN. It can be seen that a large majority of the work focussed on the SNN aspect (details of papers used in the survey are available at \cite{Survey_excel}). Most of these work focus on new materials or device fabrication and then present SNN simulations using the novel device properties bypassing the circuit level. Hence, we also decided to create a survey of ML accelerator integrated circuits (IC) published in IEEE ISSCC and IEEE SOVC conferences inspired by the ADC survey\cite{Survey_adc}. In addition, we also considered papers published in the IEEE Journal of Solid State Circuits (JSSC). Figure \ref{fig:venn_asic} plots the result of categorizing all ML accelerators adopting non von-Neumann architecture published between 2017-2019 (details in \cite{Survey_excel}). Surprisingly, it can be seen that only 5 papers have used the term ``neuromorphic" to describe their work! This clearly shows a stark difference in terminology used across different research communities.
\begin{figure}[h!]
\centering
\includegraphics[scale=0.225]{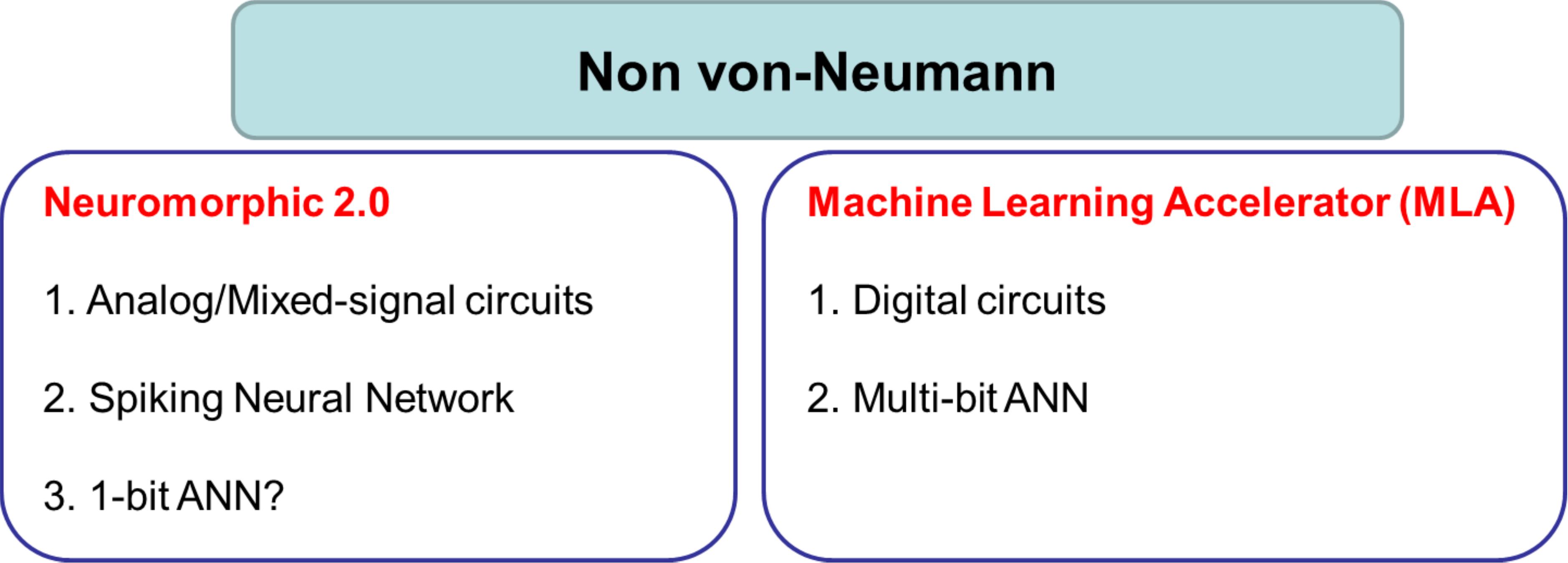}
\caption{A new taxonomy that has non von-Neumann architecture as the overarching topic with neuromorphic v2.0 and ML accelerators as two sub-topics under it.}
\label{fig:new_taxonomy}
\end{figure}
\begin{figure*}[ht!]
\centering
\includegraphics[scale=0.3]{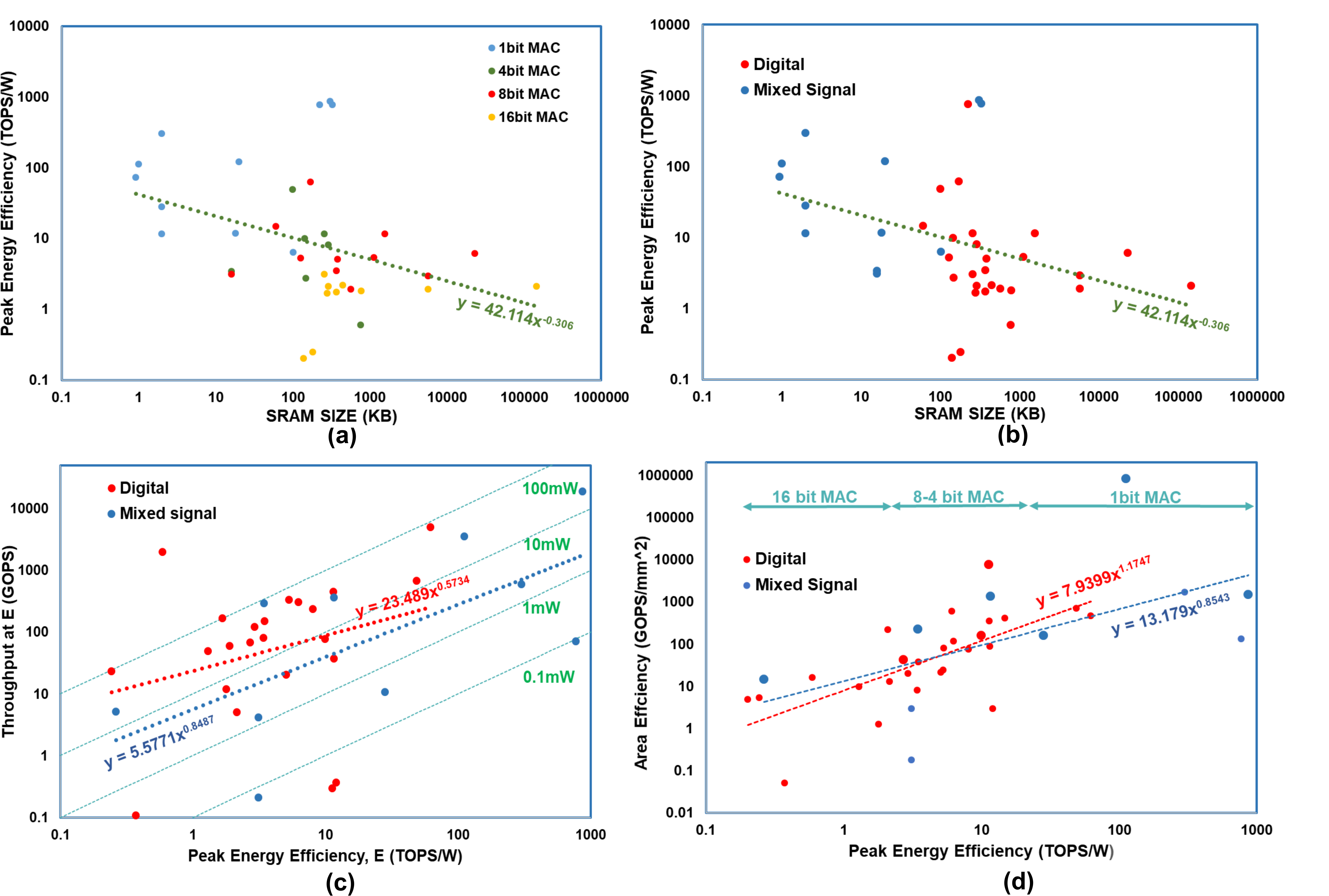}
\caption{Machine learning hardware trends: Peak energy efficiency in TOPS/W plotted against memory size and categorized by (a) bit-width of datapath and (b) digital vs mixed-signal analog approaches. (c) Throughput at peak energy efficiency and (d) Area efficiency plotted against peak energy efficiency for recent ASIC implementations reported over 2017-2019 in ISSCC, SOVC, and JSSC. The larger dots in (d) indicate ASIC area without pad.
%~\cite{A_29TOPS}~-\cite{sovc_2019_6}.
}
\label{fig:asic_trends}
\end{figure*}
This leads us to propose a new taxonomy for neuromorphic systems as shown in Figure \ref{fig:new_taxonomy}. It is possibly better to use the term non von-Neumann architecture as the overarching topic. Under its ambit, neuromorphic v2.0 can refer to systems using analog or mixed-signal circuits, implementing SNN algorithms or the extremely quantized version (1-bit) of ANNs. On the other hand, ML accelerators can refer to digital circuits with non von-Neumann architecture implementing multi-bit ANN. With this in mind, we look at some recent performance trends in ML accelerators using non von-Neumann architectures that were reviewed in Figure \ref{fig:venn_asic}.

\section{Trends in Machine Learning Hardware}
\label{sec:trends}

There are several important metrics to quantify the performance of ML accelerators such as energy efficiency, throughput and area efficiency. To identify some trends, we plot several combinations of these quantities in Figure \ref{fig:asic_trends}. 

First, we expect bigger chips to have lower energy efficiency in general due to cost of moving data around large areas that dissipates more energy charging and discharging interconnects. Since the area of these ICs are dominated by the static random access memory (SRAM) required to store weights and activations, we use the SRAM size as a proxy for chip area. The energy efficiency in Tera operations (TOPS) per Watt are plotted against SRAM size for these designs in Fig. \ref{fig:asic_trends}(a) and (b) and indeed show an inverse relation between energy efficiency and SRAM size or chip size. Figure \ref{fig:asic_trends}(a) further uses different colours to categorize the data points according to bit width of datapath. As expected, it can be seen that the extremely quantized 1-bit designs\cite{CIFAR_10,Tile,sovc_2018_3} show best energy efficiency and are located significantly ($\approx 10X$) above  the trend line. The same data is plotted in Fig. \ref{fig:asic_trends}(b) but colour coded according to the design approach of digital versus analog mixed-signal. It is interesting to note that the mixed signal designs indeed exhibit higher energy efficiencies, but they are in general much smaller than the digital ones.  

Thus, in general we can see that the neuromorphic v2.0 principles of non von-Neumann architecture coupled with low data precision and analog computing (described earlier in Section \ref{sec:taxonomy}) do indeed provide great energy efficiencies. However, it can be seen that the energy efficiencies of pure digital approaches using only the principles of non von-Neumann architecture and low bit-width are at least much higher ($\approx 500X$) than the energy efficiency wall of $\approx 10$ GMACs/W for traditional von-Neumann processors\cite{bo_wall,roadmap}. Hence, this raises an interesting question--given the scalability, testability and ease of porting across nodes offered by digital designs, is it reasonable to expect large scale industrial adoption of analog neuromorphic designs for an extra $10X$ in energy efficiency?

Next, we analyze the trade-offs in throughput at peak energy efficiency by plotting it against peak energy efficiency in Fig. \ref{fig:asic_trends}(c). Interestingly, these two quantities are positively correlated with a majority of designs exhibiting throughput $\approx 100$ GOPS. Higher throughput would generally mean the static power is better amortized across the operations leading to higher energy efficiency. Also, in general reduced bit precision designs that increases energy efficiency would also reduce critical path delays increasing throughput. Lastly, we analyze area efficiency of the designs (measured in GOPS/mm$^2$) by plotting it against energy efficiency in Fig. \ref{fig:asic_trends}(d). Again, these two quantities show a positive correlation implying again that good design practices of reduced bit precision and analog design positively impact both the quantities. This is also clarified in Figure \ref{fig:asic_trends} by demarcating the designs according to bit precision and design styles. These plots show that apart from energy efficiency, analog mixed signal design styles also provide $\approx 10X$ improvement in throughput and area efficiency. Coupled with energy efficiency advantages, these points might be sufficient to suggest that in the longer term, there is reason for large scale interest in neuromorphic designs following the principles outlined earlier. However, all of these comparisons are not very relevant unless they can all run a common set of benchmark problems. This is discussed next in the following section. 

\section{Neuromorphic Benchmarks}
\label{sec:benchmarks}
The comparison between all the hardware designs in the earlier section are not fair unless they can all at least report performance on a minimum set of benchmark algorithms. While there are at least some common benchmarks for the ANN community such as MNIST\cite{mnist}, CIFAR\cite{cifar} and Imagenet\cite{imagenet} for image recognition, there is not much consensus about good benchmarks for neuromorphic SNN algorithms. Hence, while advocating usage of benchmarks from the ML community for neuromorphic hardware ANNs, we discuss more details about what might constitute desired criteria for SNN benchmarks. While this topic is deemed important, there has been very few dedicated efforts in this area\cite{frontiers_benchmark}. Given the role the Imagenet benchmark played in catalysing progress in ANN research, we believe it is of utmost importance that the neuromorphic community spend more effort immediately on devising good benchmarks for SNN. 

Some recent work on SNN has focussed on converting images from ANN benchmarks to spike trains and then classifying them\cite{snn_abhronil,snn_SCLIU}. While being great pieces of research, we feel that this is fundamentally not a good application for SNN since the original input signal is static and does not change with time. Instead, it might be more natural to use SNN as dynamical systems to track moving objects in video streams\cite{Jyoti_socc,Jyoti_iscas} or classify signals that vary over time such as speech\cite{Jyoti_frontiers}. With this in mind, we propose the following desired characteristics for neuromorphic benchmarks:
\begin{enumerate}
\item The signal being processed should be encoded in time naturally so that the continuous time dynamics of SNN can be more effective than ANN to process it. Signals such as speech, video, etc are good examples. From the biomedical domain, EEG signals are another good example.
\item There should be a need for real-time response of the system such as in closed-loop systems such as sensori-motor loops in robotics. The rapid response time of neuromorphic sensors and SNN processing should be useful in such cases.
\item There should be need for the system to adapt or learn frequently. This necessitates learning from few samples, a common complaint with current deep learning based ANNs that require many thousands of examples to train.
\item Ideally, the example applications should be ones that require low-power operation so that the energy efficiency of neuromorphic hardware meets an important design requirement.
\item There would potentially be different benchmarks for different scales of the problem--edge deployment (sensory information processing) or cloud based analytics (large scale search, creativity etc).
\end{enumerate}
We argue in the next section that brain-machine interfaces provide a benchmark application that meets all of the above criteria.
\begin{figure}[h!]
\centering
\includegraphics[scale=0.4]{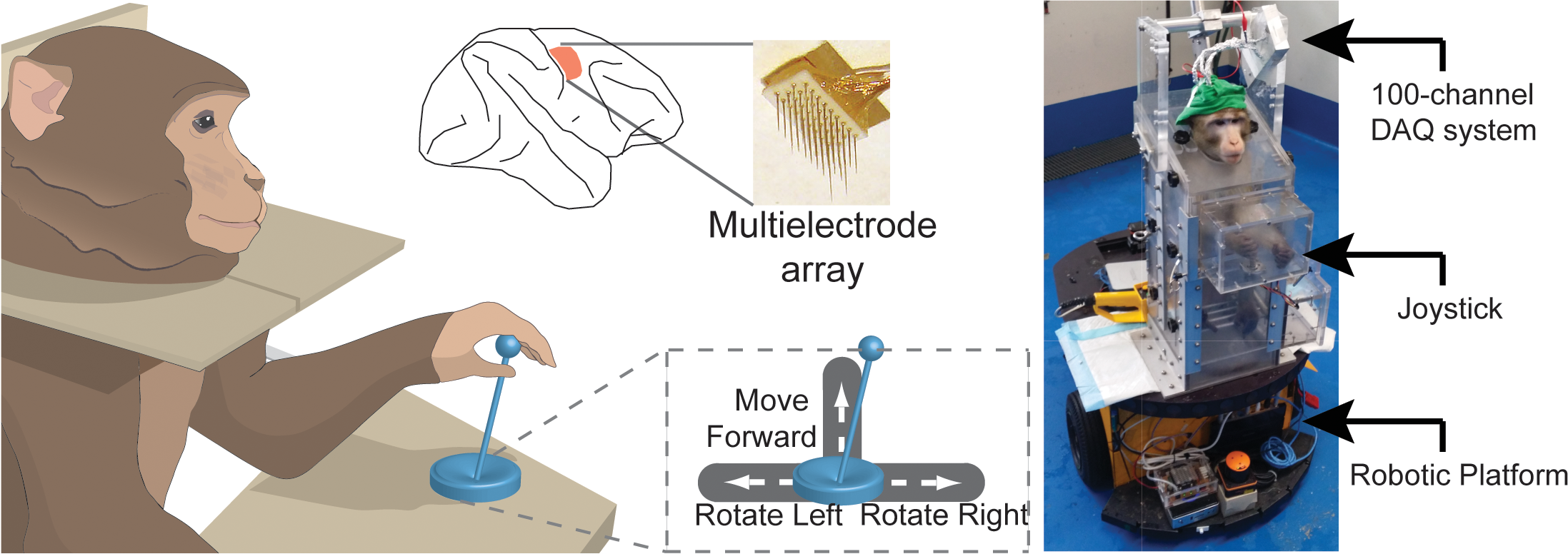}
\caption{Example of a BMI experimental setup where the NHP is using his thoughts to move a wheelchair (adopted from \cite{Libedinsky2016b} under CC-BY license). The decoder to convert brain signals to a command provides ideal opportunity for low-power, real-time neuromorphic machine learners.}
\label{fig:bmi}
\end{figure}
\section{Brain-Machine Interfaces}
\label{sec:bmi}
The aim of intra-cortical Brain Machine Interfaces (iBMIs) is to substantially improve the lives of patients afflicted by spinal cord injury or debilitating neurodegenerative disorders such as tetraplegia, amyotrophic lateral sclerosis. These systems take neural activity as an input and drive effectors such as a computer cursor \cite{Pandarinath2017}, wheelchair \cite{Libedinsky2016b} and prosthetic \cite{Collinger2013}, paralysed \cite{Ajiboye2017} limbs  for the purposes of communication, locomotion and artificial hand control respectively. While early work focussed on non-invasive EEG based systems, invasive neural interfaces are needed for fine grained motor control as well as for advancing fundamental knowledge about the brain due to higher signal quality obtainable. Figure \ref{fig:bmi} demonstrates a typical experimental setup involving a non-human primate (NHP) where an implanted micro-electrode array is interfaced with amplifiers to readout neural activity at the level of single cells\cite{Libedinsky2016b}. This neural data is collected while the primate is doing different types of tasks according to a given cue (typically visual). Based on the recorded data, a machine learner or decoder is trained to convert the neural recording to an action that affects the physical world and provides feedback to the NHP (again typically visual feedback is used). We argue that a decoder in iBMI satisfies all the conditions required for a neuromorphic system described in Section \ref{sec:benchmarks} as explained below:
\begin{enumerate}
    \item Neural data recorded from the brain are indeed a streaming signal arriving continuously over time. Further, the data are naturally in the form of spikes avoiding the question for the need of spike conversion and how to do it.
    \item Due to the visual feedback provided to the NHP, decoding has to be done in real-time. In this case, typical update frequencies of $10$ Hz are used\cite{shoeb_ner}.
    \item There is a need to frequently adapt the weights of the decoder since the neural data is non-stationary\cite{shoeb_biocas}. The statistics can change due to micro-motion of the electrode or scar tissue formation.
    \item The decoder must consume very little energy to prolong the battery life of the system\cite{shoeb_decoder}. If included within the implant, its area must be very small as well.
\end{enumerate}

There has been some initial work on neuromorphic decoders\cite{rahul_decoder,kwabena_decoder,giacomo_decoder,yi_decoder,shoeb_ner}. While \cite{rahul_decoder,kwabena_decoder} performed software simulations, \cite{giacomo_decoder,yi_decoder,shoeb_ner} have shown results from custom low-power neuromorphic ICs. Further, closed-loop decoding results from NHP have so far been demonstrated only in \cite{shoeb_ner}. One of the issues behind lack of results in this domain is the difficulty and cost of creating a NHP based experiment. Open-source datasets are just beginning to be available in this field\cite{Glaser2018PopulationCO,o_doherty_joseph_e_2018_1163026}. While these definitely will provide a good starting point, they cannot be used to simulate closed-loop settings. We envision that setting up AI based models to mimic closed-loop BMI experimental settings could be a good research direction for this area.

\section*{Conclusion}
%\section*{Acknowledgment}
In this paper, we reviewed the recent trend in papers published on the topic of neuromorphic engineering or computing and showed that the connotation of the term has broadened beyond its original definition of brain-inspired analog computing. Neuromorphic v2.0, as we call it in this paper, includes the concept of non von-Neumann and low precision digital computing from computer architecture and spiking neural networks from the computer science and algorithm community. However, there are differences in the way different scientific communities have used the term and a potential better taxonomy is to consider non von-Neumann computing as an umbrella under which a sub-concept is neuromorphic computing. Trends in recently published ML accelerator ICs indeed show that using the above neuromorphic concepts lead to $\approx 10X$ benefit in energy efficiency, area efficiency and throughput over digital non von-Neumann architectures. We also pointed out the need for benchmarks in SNN research and suggested some potential characteristics of such benchmarks. Finally, we pointed out that brain-machine interfaces (BMI) have all these desired characteristics of real-time response, processing time varying signals, need for quick re-training as well as strict requirement for low-power dissipation. We envision generation of BMI based benchmarks in the future for testing and standardization of different neuromorphic systems. 

\bibliographystyle{IEEEtran}
\bibliography{Asilomar.bib}

\end{document}